\documentclass[twoside]{article}
\usepackage{amsfonts,bezier}

\newcommand{\bls}[1]{\renewcommand{\baselinestretch}{#1}}
\def\noi{\noindent}

\makeatletter
\renewcommand{\section}{\@startsection{section}{1}{0pt}%
        {-3.5ex plus -1ex minus -.2ex}{2.3ex plus .2ex}%
        {\large\bf\protect\raggedright}}

\renewcommand{\subsection}{\@startsection{subsection}{2}{0pt}%
        {-3ex plus -1ex minus -.2ex}{1.4ex plus .2ex}%
        {\normalsize\bf\protect\raggedright}}

\renewcommand{\thesubsubsection}%
        {\arabic{section}.\arabic{subsection}.\arabic{subsubsection}.}
\newcommand{\para}{\@startsection{paragraph}{4}{0pt}%
        {1.5ex plus -.5ex minus -.2ex}{-1em}{\normalsize\bf}}

\renewcommand{\@oddhead}{\raisebox{0pt}[\headheight][0pt]{%
   \vbox{\hbox to\textwidth{\rightmark \hfil \rm \thepage \strut}\hrule}}}
\renewcommand{\@evenhead}{\raisebox{0pt}[\headheight][0pt]{%
   \vbox{\hbox to\textwidth{\thepage \hfil \leftmark \strut}\hrule}}}

\newcommand{\heads}[2]{\markboth{\protect\small\it #1}{\protect\small\it #2}}

\makeatother

\newcommand{\Title}[1]{\noi {\Large #1} \\}
\newcommand{\Author}[2]{\noi{\large\bf #1}\\[2ex]\noindent{\it #2}\\}

\newcommand{\Abstract}[1]{\vskip 2mm \begin{center}
        \parbox{16.4cm}{\small\noi #1} \end{center}\medskip}

\newcommand{\Ref}[1]{Ref.\,\cite{#1}}

\def\nqq{\hspace*{-2em}}
\def\nhq{\hspace*{-0.5em}}

\def\para{\paragraph}
\newcommand{\Theorem}[2]{\medskip\noi {\bf #1. \ }{\it #2}\medskip}

\newcommand{\Picture}[3]{
	\begin{figure} 	\centering \unitlength=1mm
	\begin{picture}(84,#1)
		\put(0,0){\line(0,1){#1}}            
		\put(0,0){\line(1,0){84}}
		\put(84,0){\line(0,1){#1}}
		\put(0,#1){\line(1,0){84}}
	\put(0,0){#2}                       \end{picture}
        \caption{\protect\small #3}  \smallskip \hrule \end{figure} }

\def\eq{Eq.\,}
\def\eqs{Eqs.\,}
\def\beq{\begin{equation}}
\def\eeq{\end{equation}}
\def\bear{\begin{eqnarray}}
\def\al{&\nhq}
\def\lal{&&\nqq {}}               
\def\bearr{\begin{eqnarray} \lal}
\def\ear{\end{eqnarray}}
\def\earn{\nonumber \end{eqnarray}}
\def\dst{\displaystyle}
\def\tst{\textstyle}

\def\yy{\\[5pt] {}}

\def\eql{\al =\al}

\def\d{\partial}

\def\const{{\rm const}}
\def\Half{{\dst\frac{1}{2}}}
\def\half{{\tst\frac{1}{2}}}

\def\DAL{\mathop{\raisebox{3.5pt}{\large\fbox{}}}\nolimits}

\def\Jl#1#2{{\it #1\/} {\bf #2},\ }

\def\CQG#1 {\Jl{Clas. Qu. Grav.}{#1}}
\def\DAN#1 {\Jl{Dokl. AN SSSR}{#1}}
\def\GC#1 {\Jl{Grav. \& Cosmol.}{#1}}
\def\GRG#1 {\Jl{Gen. Rel. Grav.}{#1}}
\def\JETF#1 {\Jl{Zh. Eksp. Teor. Fiz.}{#1}}
\def\JMP#1 {\Jl{J. Math. Phys.}{#1}}
\def\NP#1 {\Jl{Nucl. Phys.}{#1}}
\def\PLA#1 {\Jl{Phys. Lett.}{#1A}}
\def\PLB#1 {\Jl{Phys. Lett.}{#1B}}
\def\PRD#1 {\Jl{Phys. Rev.}{D\ #1}}
\def\PRL#1 {\Jl{Phys. Rev. Lett.}{#1}}

\def\GR{general relativity}

\def\ssph{static, spherically symmetric}
\def\fig{Fig.\,}

\def\MN{^{\mu\nu}}
\def\mN{_\mu^\nu}

\heads{K.A. Bronnikov}
{Spherically symmetric false vacuum: no-go theorems and global structure}

\bls{0.97}
\begin{document}
\twocolumn[
\thispagestyle{empty}

\rightline{\large\bf gr-qc/0104092}
\bigskip

\Title {\bf Spherically symmetric false vacuum: \yy
	 no-go theorems and global structure}

\Author{K.A. Bronnikov}
{Centre for Gravitation and Fundam. Metrology, VNIIMS,
        3-1 M. Ulyanovoy St., Moscow 117313, Russia;\\
Institute of Gravitation and Cosmology, PFUR,
        6 Miklukho-Maklaya St., Moscow 117198, Russia}

\Abstract
     {We enumerate all possible types of spacetime causal structures
     that can appear in \ssph\ configurations of a self-gravitating,
     real, nonlinear, minimally coupled scalar field $\varphi$ in general
     relativity, with an arbitrary potential $V(\varphi)$, not necessarily
     positive-definite. It is shown that a variable scalar field adds
     nothing to the list of possible structures with a constant $\varphi$
     field, namely, Minkowski (or AdS), Schwarzschild, de Sitter and
     Schwarzschild --- de Sitter.  It follows, in particular, that, whatever
     is $V(\varphi)$, this theory does not admit regular black holes with
     flat or AdS asymptotics. It is concluded that the only possible
     globally regular, asymptotically flat solutions are solitons with a
     regular center, without horizons and with at least partly negative
     potentials $V(\varphi)$. Extension of the results to more general field
     models is discussed.
     }

] 

\section {Introduction}

    The attractive idea of replacing the black hole (BH) singularities by
    nonsingular vacuum cores traces back to the papers of the 60s by Gliner
    \cite{glin} and Bardeen \cite{bard} but remains in the scope of modern
    studies. Possible manifestations of regular BHs vary from fundamental
    particles to largest astrophysical objects and created universes ---
    for a review see \Ref{dym00}.

    Regular BHs with phenomenological sources have been discussed in, e.g.,
    Refs.\,[2, 4--7]. A class of non\-li\-near electrodynamics Lagrangians
    leading to regular BHs with magnetic field sources has been found in
    \cite{br-prd}.

    A natural question is whether or not a regular BH can be obtained as
    a false vacuum configuration with a nonlinear scalar field in \GR, i.e.,
    from the equations of motion due to the action
\beq                                                            \label{act}
	S = \int d^4 x \,\sqrt{-g} [R + (\d\varphi)^2 -2V(\varphi)]
\eeq
    where $R$ is the scalar curvature,
    $(\d\varphi)^2 = g\MN \d_\mu\varphi\d_\nu\varphi$ and $V(\varphi)$ is a
    potential.  This action, with many particular forms of $V(\varphi)$, has
    been vastly used to describe the vacuum (sometimes interpreted as a
    variable cosmological ``constant'') in inflationary cosmology, for the
    description of growing vacuum bubbles, etc.

    An attempt to construct a regular false vacuum BH was made in \Ref
    {hoso1}, with a potential having two slightly different minima,
    $V(\varphi_1) > V(\varphi_2) =0$, the Schwarzschild metric and $\varphi
    \equiv \varphi_2$  outside the horizon, the de Sitter metric and
    $\varphi \equiv \varphi_1$ inside the horizon. It was claimed that a
    reasonable matching of the solutions was possible on the horizon despite
    a finite jump of $\varphi$.  Gal'tsov and Lemos \cite{GaLem} showed that
    the piecewise solution of \Ref{hoso1} cannot be described in terms of
    distributions and requires a singular matter source on the horizon.
    They proved \cite{GaLem} that asymptotically flat regular BH solutions
    are absent in theory (\ref{act}) with any nonnegative potential
    $V(\varphi)$ (the no-go theorem). For the region outside the horizon,
    the only asymptotically flat BH solution is Schwarzschild, as follows
    from the well-known no-hair theorems (see \Ref {bek98} for a recent
    review).

    Less is known when the asymptotic flatness and/or $V\geq 0$ assumptions
    are abandoned. Meanwhile, negative potential energy densities, in
    particular, the cosmological constant $V=\Lambda < 0$ giving rise to the
    anti--de Sitter (AdS) solution, do not lead to catastrophes (if
    restricted below), are often treated in various contexts and readily
    appear from quantum effects like vacuum polarization. Systems with an
    AdS rather than flat asymptotic cause great interest in connection with
    the AdS/CFT correspondence \cite{mald}. BHs with less symmetric
    asymptotics were also considered \cite{mann95}.

    We will study the possible global behavior of \ssph\ solutions in
    theory (\ref{act}) with arbitrary $V(\varphi)$ and arbitrary
    asymptotics. It happens that the field equations leave a very
    narrow spectrum of opportunities. According to Theorem 2 to be proved
    here, the set of causal structures is the same as known for constant
    $\varphi$:  Minkowski (or AdS), Schwarzschild, de Sitter, and
    Schwarzschild --- de Sitter (not to be confused with the de Sitter ---
    Schwarzschild structure containing a de Sitter core, discussed in
    \cite{dym00,dym92,kao00}).

    A conclusion much stronger than in \Ref {GaLem}, namely, the absence of
    regular BH solutions for {\it any\/} $V(\varphi)$ and {\it any\/}
    asymptotic, then simply follows as a corollary.

    The only possible singularity-free solutions are either de Sitter-like,
    with a single ``cosmological'' horizon (such an example was recently
    described by Hosotani \cite{hoso2}), or solutions without horizons,
    including asymptotically flat ones. The latter are, however, impossible
    if $V \geq 0$, as follows from a simple theorem proved in the manner of
    the no-hair theorems.

    The conclusions obtained here can be more or less easily extended to
    other field models, as is pointed out in the last section.

    In what follows, all statements apply to \ssph\ configurations,
    and all relevant functions are assumed to be sufficiently smooth, unless
    otherwise indicated. The symbol $\DAL$ marks the end of a proof.

\section {Field equations}

    The field equations due to (\ref{act}) are
\bear
     \nabla^\alpha \nabla_\alpha \varphi + V_\varphi \eql 0,     \label{SE}
\\
    R\mN -\half \delta\mN\, R + T\mN \eql 0,                     \label{EE}
\ear
    where $V_\varphi \equiv dV/d\varphi$, $R\mN$ is the Ricci tensor and
    $T\mN$ is the energy-momentum tensor of the $\varphi$ field:
\beq
    T\mN = \varphi_{,\mu}\varphi^{,\nu}                          \label{EMT}
        	- \half \delta\mN (\d\varphi)^2 + \delta\mN V (\varphi).
\eeq

    For a \ssph\ configuration, the metric can be written in the form
\beq                                                             \label{ds}
    ds^2 = A(\rho) dt^2 - \frac{d\rho^2}{A(\rho)} - r^2(\rho)
    			(d\theta^2 + \sin^2 \theta d\phi^2),
\eeq
    and $\varphi=\varphi(\rho)$. (We choose the coordinate gauge
    $g_{tt}g_{\rho\rho} = -1$ and suppose that large radii $r$ correspond to
    large $\rho$.) \eq (\ref{SE}) and certain combinations of
    \eqs (\ref{EE}) lead to
\bear
       (Ar^2 \varphi')' \eql r^2 V_\varphi;                   \label{phi}
\\
       	      (A'r^2)' \eql - r^2 V.                          \label{00}
\\
    	      2 r''/r \eql -{\varphi'}^2 ;                     \label{01}
\\
       A (r^2)'' - r^2 A'' \eql 2 ;                           \label{02}
\\
     A' rr' + A{r'}^2 -1 \eql \half Ar^2{\varphi'}^2 - r^2 V, \label{int}
\ear
    where the prime denotes $d/d\rho$. Only three of these five equations
    are independent: the scalar equation (\ref{phi}) follows from
    the Einstein equations, while \eq (\ref{int}) is a first integral of the
    others. Given a potential $V(\varphi)$, this is a determined set of
    equations for the unknowns $r,\ A,\ \varphi$.

    \eq(\ref{02}) can be integrated giving
\bear
    \biggl(\frac{A}{r^2}\biggr)' = -\frac{2(\rho-\rho_0)}{r^4},  \label{A'}
\ear
    where $\rho_0$ is an integration constant.

\section{No-go theorems and other observations}

    Our interest will be in the generic global behavior of the solutions
    and the existence of BHs and globally regular configurations,
    in particular, regular BHs.

    Let us first make sure that (unless the potential $V$ is singular at
    some $\rho$) the full range of the $\rho$ coordinate covers all values
    of $r$, from the center ($\rho=\rho_c$, $r(\rho_c)=0$), regular or
    singular, to infinity. To do that, we will rule out such nonsingular
    configurations as wormholes, horns and flux tubes.

    By definition, a (traversable, Lorentzian) {\it wormhole\/} with the
    metric (\ref{ds}) has two asymptotics at which $r\to \infty$, hence the
    function $r(\rho)$ has at least one regular minimum. A {\it horn\/} is a
    region of space where, as $\rho$ tends to some value $\rho^*$,
    $r(\rho)\ne \const$ and $g_{tt} = A$ have finite limits while the length
    integral $l = \int d\rho/A $ diverges. In other words, a horn is an
    infinitely long 3-dimensional ``tube'' of finite radius, where the clock
    rate remains finite. Such ``horned particles'' were, in particular,
    discussed as possible remnants of black hole evaporation \cite{banks}.
    Lastly, a {\it flux tube\/} is a configuration with $r = \const$.

\Theorem{Theorem 1}
    {The field equations due to (\ref{act}) do not admit
    (i) solutions where the function $r(\rho)$ has a regular minimum,
    (ii) solutions describing a horn, and
    (iii) flux-tube solutions with $\varphi\ne\const$.
    }

\noi{\bf Proof.} \
    Since $r(\rho) \geq 0$ by its geometric meaning, \eq (\ref{01}) gives
    $r''\leq 0$, which rules out regular minima. The same equation leads to
    $\varphi =\const$ as soon as $r = \const$. Thus items (i) and (iii) have
    been proved.

    Suppose now that there is a horn. Then, by definition, $A$
    has a finite limit whereas $l\to \infty$ as $\rho\to \rho^*$. This is
    only possible if $\rho^* = \pm \infty$. Under these circumstances,
    the left-hand side of \eq (\ref{A'}) vanishes at the ``horn end'',
    $\rho\to \rho^* = \pm \infty$, whereas its right-hand side tends to
    infinity. This contradiction proves item (ii).
$\DAL$

\medskip
    Due to the local nature of the proof, Theorem 1 rules out wormholes
    or horns with any large $r$ behavior --- flat, de Sitter or any other.
    Moreover, since $r'>0$ at large $\rho$, the function $r(\rho)$ is
    monotonic in the whole range.

    Let us now address to the causal structure of the solutions,
    determined by the disposition of static ($A>0)$ and nonstatic ($A<0)$
    regions of spacetime (also labeled R and T regions, respectively).
    This relationship is unambiguous in the sense that a particular
    disposition of regions leads to a certain Penrose-Carter diagram
    \cite{walker,br79}. The latter may be further complicated by
    identification of isometric surfaces, if any, and by branching that
    leads to structures like Riemann surfaces \cite{cold,br79}.

    Horizons that separate the regions are regular spheres of nonzero
    radius, corresponding to zeros of the function $A(\rho)$. Such
    zeros, if any, are regular points of \eqs (\ref{phi})--(\ref{int}) due
    to our choice of the coordinates. Moreover, near a horizon, $\rho$
    varies (up to a positive constant factor) like manifestly well-behaved
    Kruskal-like coordinates used for an analytic continuation of the metric
    \cite{cold}. Therefore one can jointly consider regions on both sides
    of a horizon.

    A horizon is {\it simple\/} or {\it multiple\/} according to whe\-ther
    the zero of $A(\rho)$ is simple or multiple.
    A simple or, in general, odd-order horizon separates a static region
    from a nonstatic one (as, e.g., the Schwarzschild horizon).
    Even-order horizons separate regions of the same nature (as the
    double horizon in the extreme Reissner-Nordstr\"om metric).

    The following theorem severely restricts the set of possible structures.

\Picture{70}
{       \unitlength .45mm
\special{em:linewidth 0.4pt}
\linethickness{0.4pt}
\begin{picture}(149.00,136.00)
\put(15.00,65.00){\vector(1,0){132.00}}
\bezier{288}(27.00,126.00)(36.00,95.00)(75.00,86.00)
\bezier{576}(30.00,132.00)(45.00,94.00)(148.00,84.00)
\put(147.00,56.00){\makebox(0,0)[lb]{$\rho$}}
\put(32.00,144.00){\makebox(0,0)[rc]{$B=A/r^2$}}
\put(52.00,11.00){\makebox(0,0)[cc]{1}}
\put(43.00,27.00){\makebox(0,0)[cc]{2}}
\put(37.00,53.00){\makebox(0,0)[cc]{3}}
\put(36.00,101.00){\makebox(0,0)[cc]{4}}
\put(37.00,127.00){\makebox(0,0)[cc]{5}}
\put(20.00,13.00){\vector(0,1){124.00}}
\bezier{260}(75.00,86.00)(106.00,78.00)(130.00,55.00)
\bezier{80}(130.00,55.00)(140.00,47.00)(147.00,46.00)
\bezier{604}(28.00,20.00)(57.00,109.00)(113.00,96.00)
\bezier{148}(113.00,96.00)(137.00,90.00)(149.00,90.00)
\bezier{780}(34.00,11.00)(65.00,118.00)(118.00,53.00)
\bezier{144}(119.00,52.00)(132.00,39.00)(149.00,37.00)
\bezier{512}(47.00,10.00)(69.00,88.00)(102.00,55.00)
\bezier{224}(103.00,54.00)(127.00,30.00)(149.00,29.00)
\end{picture}
}
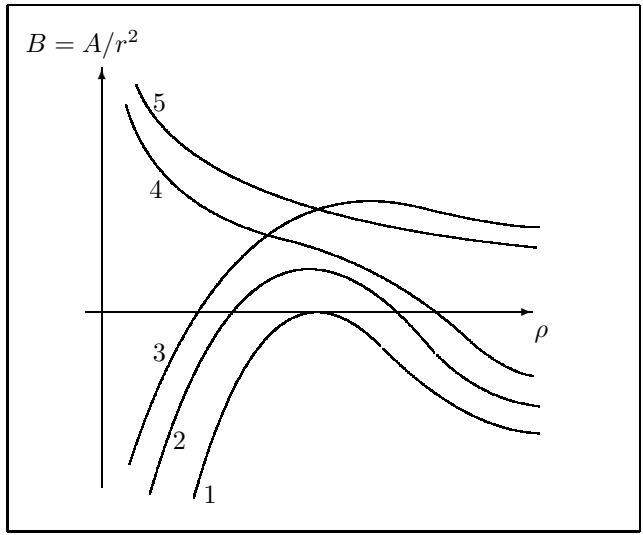
{Qualitative behaviors of the function $B(\rho)$ determining the
causal structure of spacetime: 1 --- no static region; 2 --- Schwarzschild
-- de Sitter; 3 --- Schwarzschild; 4 --- de Sitter; 5 --- Minkowski/AdS}

\Theorem{Theorem 2} {Consider solutions of the theory (\ref{act}) with
	the metric (\ref{ds}) and $\varphi=\varphi(\rho)$. Let there be a
	static region $a < \rho < b \leq \infty$. Then:
\begin{description}\itemsep -2pt
\item [(i)]
	all horizons are simple;
\item [(ii)]
	no horizons exist at $\rho < a$ and at $\rho > b$.
\end{description}
}

\noi{\bf Proof.}
    Let $\rho=h$ be a horizon: $A(h)=0$.
    It follows from \eq (\ref{02}) that $A''(h) = -2/r_h^2 < 0$.
    Therefore $h$ cannot be a zero of $A(\rho)$ of order higher than two.
    Consider the function $B(\rho) = A/r^2$. A horizon is
    also a zero of $B(\rho)$ of the same multiplicity as that of $A$.
    If it is double, $A'(h) = B'(h)=0$, then by (\ref{A'}) $h=\rho_0$,
    so that $B' > 0$ at $\rho < h$ and $B'<0$ at $\rho>0$. Thus $B < 0$ for
    all $\rho\ne h$, and the spacetime has no static region (curve 1
    in \fig 1). So item (i) is proved.

    Consider now the boundary $\rho=a$ of the static region. If $r(a)\ne 0$,
    then it is a horizon, $A(a)=B(a)=0$. (One cannot have $B(a)=\infty$
    since by (\ref{A'}) $|B'(a)| < \infty$.) By item (i), the horizon is
    simple, and $B'(a) >0$, therefore in (\ref{A'}) we have $a < \rho_0$
    whence it follows that $B' > 0$ everywhere to the left of $\rho = a$:
    $B(\rho)$ is an increasing function and cannot return to zero, ruling
    out horizons at $\rho <a$ (see curves 2 and 3 in \fig 1).

    If $b <\infty$, horizons at $\rho > b$ are ruled out in a similar manner.
$\DAL$

\medskip
    According to Theorem 2, the list of possible global
    structures is the same as the one for constant $\varphi$:
\begin{description}          \itemsep -2pt
\item[{\rm [TR]:}]
	Schwarzschild (curve 3 in \fig 1),
\item[{\rm [RT]:}]
	de Sitter (curve 4),
\item[{\rm [R]:}] \
	Minkowski or AdS (curve 5),
\item[{\rm [TRT]:}]
	Schwarzschild -- de Sitter (curve 2), and
\item[{\rm [TT], [T]:}]
	spacetimes without static regions (curve 1 and still below).
\end{description}
    The R and T letters in brackets show the sequence of static and
    nonstatic regions, ordered from center to infinity. The center is
    generically singular. The only possible nonsingular solutions have
    either Minkowski/AdS or de Sitter structures, and, in particular,
    solitonlike asymptotically flat solutions are not excluded.

\Theorem{Corollary}
    {The theory (\ref{act}) does not admit \ssph, regular BHs.}

    Indeed, such a BH, with any large $r$ behavior, must have static
    regions at small and large $r$, separated by at least two simple or
    one double horizon (in the above notation, the structure must be [RTR]
    or [RR] or more complex).  This is impossible according to Theorem 2.

    The above two theorems did not use any assumptions on the asymptotic
    behavior of the solutions or the shape and even sign of the potential.
    Let us now mention some more specific but important results.
    One of them is the well-known no-hair theorem, first proved by
    Bekenstein \cite{bek72} for $V(\varphi)$ without local maxima and
    later extended to any $V \geq 0$ and some more general Lagrangians
    (see, e.g., \Ref{bek98} for proofs and references):

\Theorem{Theorem 3}
    {Suppose $V \geq 0$. Then the only asymptotically flat BH solution to
    \eqs (\ref{phi})--(\ref{int}) in the range $(h, \infty)$ (where $\rho=h$
    is the event horizon) comprises the Schwarzschild metric, $\varphi
    =\const$ and $V\equiv 0$.
    }

    Another restriction \cite{brsh} concerns the properties of globally
    regular configurations and can be called the generalized Rosen theorem
    (G.  Rosen \cite{rosen} studied similar restrictions for nonlinear
    fields in flat spacetime):

\Theorem{Theorem 4}
    {An asymptotically flat solution with a positive mass $M$ and a regular
    center is impossible with $V(\varphi)\geq 0$.
    }

    Here is a proof slightly different from the one given in
    Ref.\,\cite{brsh}.  Integrate \eq (\ref{00}) from the center
    ($\rho=\rho_c$) to infinity:
\beq
    A'r^2 \Bigl|^\infty_{\rho_c} =                           \label{int00}
     			-2 \int_{\rho_c}^{\infty} r^2\, V\, d\rho.
\eeq
    In an asymptotically flat metric, $A(\rho)$ behaves at large $\rho$ as
    $1 - 2M/\rho^2$, where $M$ is the Schwarzschild mass in geometric units,
    and $r=\rho + O(1)$, therefore the upper limit of $A'r^2$ equals $2M$.
    At a regular center $r = 0$ and, as is easily verified, $A'=0$, so the
    lower limit is zero. Consequently,
\beq
    M = - \int_{\rho_c}^{\infty} r^2\, V\, d\rho.              \label{V-}
\eeq
    Thus $M > 0$ requires that $V(\varphi)$ should be at least partly
    negative.
$\DAL$

\medskip
    This simple theorem, proved previously for particlelike solutions
    without horizons, equally applies to regular BHs, if any.  Hence,
    independently of Theorem 2, it also rules out regular BHs with $V\geq
    0$. Theorem 2 tells us more: even negative potentials do not create
    such BHs.

\section{Generalizations}

    One can notice that Theorem 1 actually rests on the validity of the null
    energy condition which, with the metric (\ref{ds}), reads $T^t_t -
    T^\rho_\rho \geq 0$. This, via the appropriate Einstein equation,
    leads to $r'' \leq 0$.  In turn, Theorem 2 rests on \eq (\ref{02})
    following from
\beq
        T^t_t = T^\theta_\theta.                               \label{T02}
\eeq
    Thus both theorems hold for all kinds of matter whose
    energy-momentum tensors satisfy these two conditions.

    Consider, for instance, the following action, more general than
    (\ref{act}):
\beq                                                           \label{act'}
	S = \int d^4 x \,\sqrt{-g}\ [R + F(I, \varphi)]
\eeq
    where $I = (\d\varphi)^2$ and $F(I, \varphi)$ is an arbitrary function.
    The scalar field energy-momentum tensor is
\beq
	T\mN = \frac{\d F}{\d I}
         	\varphi_{,\mu}\varphi^{,\nu}                   \label{EMT'}
        	               - \Half \delta\mN F (\varphi).
\eeq
    In the \ssph\ case, \eq (\ref{T02}) holds automatically due to
    $\varphi=\varphi(\rho)$, while the null energy condition holds
    as long as $\d F/\d I \geq 0$, which actually means that the kinetic
    energy is nonnegative. Under this condition, both Theorems 1 and 2 are
    valid for the theory (\ref{act'}). Otherwise Theorem 2 alone holds; it
    still correctly describes the $\rho$ dependence of $A$ and consequently
    the possible horizons disposition, but $r(\rho)$ is not necessarily
    monotonic, so that wormholes and horns are not forbidden.

    An extension of the present results to higher dimensions, with
    coordinate spheres ${\mathbb S}^{D-2}$ instead of ${\mathbb S}^2$, is
    straightforward. Other extensions, which need investigation, concern
    theories connected with (\ref{act}) and (\ref{act'}) by
    $\varphi$-dependent conformal transformations, such as theories with
    nonminimally coupled scalar fields (e.g., scalar-tensor theories) and
    nonlinear gravity (e.g., with the Lagrangian function $f(R)$). I hope to
    consider them in future papers.

\subsection*{Acknowledgements}

This work was supported in part by the Russian Foundation for Basic
Research and by NATO Science Fellowship Programme grant.
I acknowledge kind hospitality at the Dept. Math. at the University of
the Aegean, Karlovassi, Samos, Greece, where part of the work was done. I am
grateful to Spiros Cotsakis, Vitaly Melnikov, George Shikin and Olaf
Lechtenfeld for helpful discussions.

\end{document}